\documentclass{article}

\usepackage{mathrsfs}
\usepackage{amsfonts}
\usepackage{amsmath}
\usepackage{amssymb}
\usepackage{graphicx}
\newtheorem{Conjecture}{Conjecture}

\title{The Duality of the Universe}
\author{Gordon McCabe}

\begin{document}

\maketitle

\begin{abstract}
It is proposed that the physical universe is an instance of a mathematical
structure which possesses a dual structure, and that this dual structure is the
collection of all possible knowledge of the physical universe. In turn, the
physical universe is then the dual space of the latter.

\end{abstract}

\section{Introduction}

The purpose of this paper is to propose that the physical universe, and all
knowledge of the physical universe, are related to each other by a mathematical
duality relationship. Whilst this proposal was inspired by the hypothesis in
Majid (2000, 2007) and Heller (2004a,b) that the physical universe is
\emph{self}-dual, the proposal made here suggests, on the contrary, that the
dual space of the physical universe possesses a distinct structure to that of
the physical universe. The idea proposed in this paper does, however, share
with Heller's work the hypothesis that the physical universe is an instance of
a mathematical structure. Equivalently, it can be asserted that the physical
universe is isomorphic to a mathematical structure (Tegmark, 1998). This
doctrine can be dubbed `universal structural realism'.

In particular, the proposals in this paper draw upon concepts from mathematical
category theory, and the notion of a mathematical dual. A category consists of
a collection of objects such that any pair of objects has a collection of
`morphisms' between them. The morphisms satisfy a binary operation called
composition, which means that you can tack one morphism onto the end of
another. In addition, each object has a morphism onto itself called the
identity morphism. For example, the category Set contains all sets as objects
and the functions between sets as morphisms; the category of topological spaces
contains all topological spaces as objects, and has continuous functions as
morphisms; and the category of smooth manifolds contains all smooth manifolds
as objects and has `smooth' (infinitely-differentiable) maps as morphisms.
However, the definition of a category does not require the morphisms to be
special types of functions, and the objects need not be special types of set.

Categories can be related by maps called `functors', which map the
objects in one category to the objects in another, and which map
the morphisms in one category to the morphisms in the other
category, in a way which preserves the composition of morphisms.
Furthermore, one can relate one functor to another by something
called a `natural transformation'. Suppose that $o$ is an object
in a category $A$, and suppose that $X$ is a functor from $A$ to
another category $B$, and $Y$ is a functor from $A$ to another
category $C$. $X$ maps $o$ to $X(o)$, an object in category $B$,
and $Y$ maps $o$ to $Y(o)$, an object in category $C$. A natural
transformation $N$ from $X$ to $Y$ defines the image of $X$ in
$Y$. Hence, a natural transformation is defined by a family of
maps, $N_o: X(o) \rightarrow Y(o)$, for all the objects $o$ in the
category $A$.

There is a special type of category called a monoidal category
$\mathscr{C}$ which is equipped with a unit object $\underline{1}$
and an associative product functor $\otimes: \mathscr{C} \times
\mathscr{C} \rightarrow \mathscr{C}$, such that $V \otimes
\underline{1} \cong V \cong \underline{1} \otimes V$, for any $V
\in Obj(\mathscr{C})$.

Given a vector space $V$ over a number field $\mathbb{F}$, the
dual space is defined to be the space $V^*$ of linear functionals
$\phi:V \rightarrow \mathbb{F}$. $V^*$ is also a vector space over
$\mathbb{F}$, hence one can take the dual of the dual $V^{**}$.
This is actually isomorphic to the original vector space, $V^{**}
\cong V$, because each element $v \in V$ defines a linear
functional $v: V^* \rightarrow \mathbb{F}$ by the stipulation that
$v(\phi) = \phi(v)$.

In the case of a commutative group $G$, the dual $\hat{G}$ is
defined to be the set of homomorphisms $\phi:G \rightarrow S^1$
into the set of complex numbers of unit modulus. This coincides
with the set of one-dimensional unitary representations of $G$,
and it transpires that in a general context, the dual space is a
space of representations.

In particular, the form of duality proposed in this paper is a
generalisation of that established by the Tannaka theorem for
non-commuatative groups. Given such a group $G$, the dual object
is the category $\Pi(G)$ of finite-dimensional unitary
representations, a monoidal category whose product is the tensor
product of representations. A representation $\phi$ of the
category $\Pi(G)$ is defined to be such that it associates with
each object $T: G \rightarrow End(V)$ in $\Pi(G)$ an endomorphism
of the target space of $T$,

$$
\phi(T) \in End(V) \;.
$$ With each $g \in G$, one can define such a representation $\phi_g$ as follows:

$$
\phi_g(T) = T(g) \in End(V) \;.
$$ The Tannaka theorem demonstrates that the map $g \mapsto \phi_g$ is an isomorphism between $G$ and $\Gamma(\Pi(G))$,
 the space of all representations of $\Pi(G)$. Hence, taking the space of all representations of $\Pi(G)$ to be the
 dual of $\Pi(G)$, the dual of the dual is isomorphic to the
 original group, $\Gamma(\Pi(G)) \cong G$.

\section{Epistemology is the dual of metaphysics}

Equipped with these concepts, we now proceed to outline the main idea of the
paper as a series of propositions. Our first proposition is that:

\begin{Conjecture}The structure of the physical universe is such that any two models
of this structure are isomorphic. Such a structure corresponds to
a category $\mathscr{C}$ containing only one object
$\mathfrak{U}$.
\end{Conjecture}

Our second proposition is that:

\begin{Conjecture}The representations of $\mathfrak{U}$ are provided by
the collection of functors $Funct(\mathscr{C})$ from $\mathscr{C}$ into other
categories. $Funct(\mathscr{C})$ is the dual space of
$\mathfrak{U}$.\end{Conjecture}

This is simply a generalisation of the familiar concept of a linear group
representation. One can think of a group $G$ as a category with only one
object, in which all the morphisms (the group elements) are isomorphisms. Under
this perspective, a linear group representation $T:G \rightarrow End(V)$ is a
functor from the single-object category into the category of vector spaces.
Such a functor maps the single object to a vector space, and maps the group
elements, (the morphisms of the single object), into the morphisms of the
vector space $V$, (the linear operators $End(V)$). Generalising this, a linear
representation of an arbitrary category is a functor from that category into
the category of vector spaces, and a general representation of an arbitrary
category is a functor from that category into some other category.

Our third proposition is:

\begin{Conjecture}Knowledge of the physical universe
corresponds to a representation of the universe, hence all possible knowledge
of the universe corresponds to the collection of all possible representations
of the universe. The collection of all possible representations of the universe
corresponds to the collection of functors $Funct(\mathscr{C})$, hence the
collection of all possible knowledge of the universe corresponds to
$Funct(\mathscr{C})$. The collection of all possible knowledge of the physical
universe is the dual space of the universe.
\end{Conjecture}

We shall elaborate on this proposition below. For now, suffice to note that
knowledge of the universe corresponds not necessarily to homomorphisms of the
structure $\mathfrak{U}$, or its substructures, but, more generally, to
functors into objects in other categories.

Our fourth proposition is that

\begin{Conjecture}The dual $\Gamma(Funct(\mathscr{C}))$ of the dual is isomorphic to
$\mathfrak{U}$.\end{Conjecture}

Given a collection of functors $Funct(\mathscr{V})$ of a category
$\mathscr{V}$, a representation of the collection of functors assigns to each
functor $T: \mathscr{V} \rightarrow \mathscr{W}$ an object in the target
category $\mathscr{W}$. Specifically, for each object $v \in Obj(\mathscr{V})$,
there is a representation $\phi_v$ on the collection of functors defined as
follows:

$$
\phi_v(T)=T(v) \in Obj(\mathscr{W}) \;.
$$ Thus, for a category $\mathscr{V}$, the space of representations of the space of representations, is isomorphic to $\mathscr{V}$.

Our fifth proposition is that

\begin{Conjecture}Knowledge of all the knowledge of the physical universe
corresponds to a representation of $Funct(\mathscr{C})$. Given that the
collection of all such representations is isomorphic to $\mathfrak{U}$, it
follows that all knowledge of all knowledge of the physical universe is
isomorphic to the physical universe itself $\mathfrak{U}$. Thus, one can
re-construct the structure of the physical universe from the collection of all
knowledge of the physical universe. The physical universe is the dual space of
the collection of all knowledge of the physical universe.\footnote{Note that
although it is postulated here that the dual space of the physical universe is
distinct from, and non-isomorphic to the physical universe itself, as Majid
(2007, p15) points out, given a structure $X$ and its dual $\hat{X}$, one can
form a self-dual structure by simply taking the cartesian product $X \times
\hat{X}$.}
\end{Conjecture}

If we take metaphysics to be the most general study of the physical world, and
epistemology to be the most general study of all knowledge of the physical
world, then epistemology is the most general study of the dual space of the
physical world. Epistemology is the most general study of the space dual to
that which metaphysics provides the most general study of. In this sense,
epistemology can be said to be the dual of metaphysics.

The third proposition, that defines knowledge to consist of mathematical
representations of the physical universe by means of functors, is sufficiently
general to encompass the various different types of cognitive representation
present within our culture. C.S Peirce proposed a tripartite division of
representational `signs' into `icons', `indices', and `symbols'. Peirce held
that icons resemble what they represent, indices are causally connected to what
they represent, and symbols are arbitrary labels for what they represent, (see
Schwartz 1995, p536-537). Each of these types of representation can be
considered to be functors. A physical object, or the state of a physical
object, can be represented by a mapping $f$ if either:

\begin{enumerate}
\item{The object/state is a structured entity $\mathscr{M}$, which is the domain of a mapping $f: \mathscr{M} \rightarrow f(\mathscr{M})$
defining the representation. The range of the mapping, $f(\mathscr{M})$, will
also be a structured entity, and the mapping $f$ will be a homomorphism with
respect to some level of structure possessed by $\mathscr{M}$ and
$f(\mathscr{M})$.}
\item{The object/state is an object $x$ in a category $\mathscr{C}$, and the
mapping $f: \mathscr{C} \rightarrow f(\mathscr{C})$ defining the representation
is a functor. As a special case, if $\mathscr{C}$ is a set $\mathscr{M}$, and
the object/state is an element $x \in \mathscr{M}$, then the mapping $f:
\mathscr{M} \rightarrow f(\mathscr{M})$ defining the representation will simply
be a map between sets.}
\end{enumerate}

In the first case, $\mathscr{M}$ and $f(\mathscr{M})$ can be treated as objects
belonging to different categories, and the mapping $f$ can be treated as the
restriction to $\mathscr{M}$ of a functor between those categories.

The first type of representational mapping corresponds mathematically to a
homomorphism. A scale-model of a Formula 1 car, or the topological map of the
London underground, exemplify this type of representation. The homomorphism
between a particular Formula 1 car and its wind-tunnel model is the restriction
of a functor between the category of all Formula 1 cars and the category of
their wind-tunnel models.

The second type of representational mapping, in contrast, doesn't correspond to
a homomorphism between a thing and its representation. In this case, the
representational functor or mapping $f: \mathscr{C} \rightarrow f(\mathscr{C})$
can be defined by either (i) an objective, causal physical process, or by (ii)
the decisions of thinking-beings.

The primary example of type-i non-homomorphic representation is the perceptual
representation of the external world by brain states. Taking the example of
visual perception, there is no homomorphism between the spatial geometry of an
individual's visual field, and the state of the neuronal network in that part
of the brain responsible for vision. Nevertheless, the correspondence between
brain states and the external world is not an arbitrary mapping, but a
correspondence defined by a causal process involving photons of light, the
human eye, the retina, and the human brain. The correspondence exists
independently of human decision-making.

The primary example of type-ii non-homomorphic representation is the
representation of a physical system provided by a digital computer simulation.
A contemporary computer represents a physical system by electronically encoding
the numerical representation provided by mathematical physics. Numbers are
represented by segments of computer memory called `words', typically consisting
of several bytes. There is no homomorphism between a number and the electronic
state of a word of computer memory; each number is merely an element in the
domain of a mapping which maps numbers to the electronic states of computer
memory. There are many ways to represent a number by the state of a word of
computer memory. Moreover, the same electronic states of computer memory can
represent things other than numbers, such as character symbols, or images and
sounds. The correspondence between numbers and states of computer memory is
dependent upon the interpretational decisions taken by the humans who program
and operate the simulation.

\section{Mental representation}

The proposal that knowledge consists of various types of mathematical
representation, incorporates the representational theory of the mind (RTM), an
approach to the mind-brain relationship which falls under the aegis of
functionalism. Let us briefly digress, then, to explain this approach to the
mind-brain relationship.

Functionalism is one possible reaction to the identity theory of the mind-brain
relationship. The identity theory claims that minds can be reduced to brains in
the sense that mental properties, states and processes can be \emph{defined} in
terms of brain properties, states and processes. Functionalism rejects this,
but endorses the weaker notion of supervenience, which holds that any change in
the higher level properties, states and processes of a composite system, must
correspond to a change in the lower level properties, states and processes. The
idea is that there can be no difference in the higher-level state of a
composite system without a difference in the lower-level state, otherwise there
would be a one-many correspondence between the lower-level states and
higher-level states. In these terms, the mind clearly supervenes upon the
brain: each brain state determines a unique mental state, and each change in
mental state requires a change of brain state. However, supervenience does not
require the higher-level description to be definable in terms of the
lower-level description. In particular, supervenience does not require the
properties of the higher-level description to be definable in terms of the
properties of the lower-level description. When supervenience is combined with
this claim of irreducibility, the resulting credo is often termed
`emergentism'. Emergent states and properties are higher level states and
properties which supervene upon lower level states and properties, but which
are not definable in terms of those lower-level properties.

Functionalism accepts that the mind and the brain exist; it accepts that the
mind cannot be identified with the brain; and it accepts that the mind
supervenes on the brain. Functionalism claims that the mind is a set of
functionalities and capabilities, at a higher level of description than the
brain. Thus, although functionalism accepts that the mind cannot be identified
with the brain, it still contends that the mind can be objectively
characterised.

The notion that the identity of the mind is defined by a set of functionalities
and capabilities, leads to the notion of substrate-independence, the claim that
the mind could supervene upon multiple substrates, of which the brain just
happens to be one example. Neurophysiology demonstrates how brain structure
supports the functionalities and capabilities of the mind, hence
neurophysiology demonstrates how the structure of the brain supports these
mental structures. However, functionalism argues that there are multiple
substrates which could support such mental structures, hence one cannot
identify the mind with the brain.

From the perspective of structural realism, functionalism holds that the mind
possesses a structure which cannot be identified with the structure of the
brain. In other words, functionalism holds that the structure of the mind is
non-isomorphic to the structure of the brain. This is consistent with the
proposal that the structure of the physical universe has a dual structure,
which is non-isomorphic to the physical universe.

Functionalism holds that a mental state has functional relationships to
perceptual stimuli, behavioural responses, and other mental states. i.e., a
mental state maps current perceptual stimuli to behavioural responses and the
next mental state. Thus, in mathematical terms, one can treat a mental state as
a function

$$I \rightarrow S \times O \;,$$ where $I$ is the set of
input states (the perceptual states), $S$ is the set of mental states, and $O$
is the set of output states, (the behavioural responses). By implication, the
set of mental states is then a set of such functions, and this set presumably
possesses some structure.

Functionalism, however, should not be conflated with `behaviourism', which
claims that mental states have no `internal' content, and are \emph{nothing
but} maps between perceptual stimuli and behavioural responses. Functionalism
claims that mental states are a lot more than such maps. We shall now consider
two specific functionalist approaches to the mind: the `top-down' approach
provided by the conjunction of the representational theory of the mind (RTM)
and the computational theory of the mind (CTM), and the `bottom-up' approach of
\emph{connectionism}. In particular, we shall attempt to highlight the
structural aspects of these approaches.

The RTM attempts to provide a functionalist account of `intentional' mental
states. These are states, such as beliefs and desires, in which the attention
of the mind is directed towards something, called the `content' of the
intentional state. The RTM claims that an intentional mental state is a type of
functional state that involves a relationship between the thinker and the
symbolic representation of something. If a thinker holds the belief that `the
cat is on the mat', then the thinker is held to be in one type of functional
relationship to a symbolic representation of `the cat is on the mat'. If a
thinker holds the desire that `the cat is on the mat', then the thinker is held
to be in a different type of functional relationship to the same symbolic
representation. Beliefs and desires differ by virtue of the fact that they
cause different succeeding mental states and behavioural responses. The RTM
considers mental processes such as thinking, reasoning and imagining to be
sequences of intentional mental states.

Many advocates of the RTM claim that the mental representations which provide
the content of beliefs, desires, and other intentional states, possess an
internal structure. They hold that this internal system of representation has a
set of symbols, a syntax, and a semantics, collectively termed the language of
thought. There are rules for composing the symbols into expressions,
propositions, and mental images, hence the content of an intentional state can
be said to possess a symbol structure, and one might call this the
infrastructure of an intentional state. The computational theory of mind (CTM)
is then the conjunction of the RTM with the claim that mental reasoning is the
formal, syntactical manipulation of such symbols (Horst 2005). Applying the
main proposal in this paper, an intentional mental state belongs to a category
of intentional states, which are related by a functor to the things they
represent.

In contrast with the CTM, connectionism does not attempt to model the mind by
ascribing a structure to intentional states at the outset. Instead,
connectionism adopts a `bottom-up' approach, modelling the mind with `neural
networks', abstractions from the network of nerve cells and synapses in the
human brain. In the connectionist approach, the foremost structures are those
possessed by the neural networks, not those possessed by sets of intentional
states, or those possessed by the symbols which purportedly compose the
contents of intentional states.

A neural network consists of a set of nodes, and a set of connections between
the nodes. The nodes in a neural network possess activation levels, the
connections between nodes possess weights, and the nodes have numerical rules
for calculating their next activation level from (i) the previous activation
level, and (ii) the weighted inputs from other nodes. The pattern of
connections and activation levels can be thought to provide a lower-level of
description than that provided by sets of intentional states and symbol
structures. However, under certain circumstances, the pattern of connections
and activation levels is deemed to provide a representation of things, hence
the connectionist models of the mind can still be subsumed under the aegis of
the RTM. It should also be noted that connectionism (arguably) contrasts with
the CTM in the sense that thinking can take place at a sub-symbolic level.

These functionalist accounts of the mind entail that the mind \emph{can} be
incorporated into structural realism. They claim that the mind possesses a
structure, albeit a higher-level structure than that of the brain. All
functionalist accounts accept that the mind supervenes on the brain, but they
reject the idea that the higher-level structure can be defined in terms of the
lower-level structure. One can accept that the mind is essentially subjective,
but one can still hold that it possesses a structure, a distinct structure from
the structure of the brain. The structure may well be a non-spatial structure,
which cannot be reduced to any of the structures which characterise the brain,
but it is, nevertheless, the structure of subjective experience. Intentional
mental states are not observable by means of the sense organs, are directly
accessible only to their owners, and do not occupy space, yet they nevertheless
possess a structure which functionalism and the RTM sets out to capture. The
proposal made in this paper entails that this structure is part of the dual
structure of the physical universe.

\end{document}